# Thin films of the spin ice compound Ho$_2$Ti$_2$O$_7$

D. P. Leusink[1], F. Coneri[1], M. Hoek[1], S. Turner[2], H. Idrissi[2], G. Van Tendeloo[2], H. Hilgenkamp[1*]

[1]Faculty of Science and Technology & MESA+ Institute for Nanotechnology, University of Twente, P.O. Box 217, 7500 AE Enschede, The Netherlands. [2]Electron Microscopy for Materials Science (EMAT), Department of Physics, University of Antwerp, Groenenborgerlaan 171, B-2020 Antwerp, Belgium.
*e-mail: h.hilgenkamp@utwente.nl

The pyrochlore compounds Ho$_2$Ti$_2$O$_7$ and Dy$_2$Ti$_2$O$_7$ show an exotic form of magnetism called the spin ice state, resulting from the interplay between geometrical frustration and ferromagnetic coupling. A fascinating feature of this state is the appearance of magnetic monopoles as emergent excitations above the degenerate ground state. Over the past years, strong effort has been devoted to the investigation of these monopoles and other properties of the spin ice state in bulk crystals. A tantalising prospect is to incorporate spin ice materials into devices for spintronics and devices that can manipulate the magnetic monopoles. This would require the availability of spin ice thin films. Here, we report the fabrication of Ho$_2$Ti$_2$O$_7$ thin films using pulsed laser deposition. These films not only show a high crystalline quality, but also exhibit the hallmarks of a spin ice: a pronounced magnetic anisotropy and an intermediate plateau in the magnetisation along the [111] crystal direction.

The magnetic Ho$^{3+}$ ions in Ho$_2$Ti$_2$O$_7$ form a lattice of corner-sharing tetrahedra in the pyrochlore structure and interact via ferromagnetic coupling[1-7]. Their magnetic moments are Ising-like due to the crystal field anisotropy and are aligned along the set of ‹111› axes. The resulting degenerate ground state of each tetrahedron has two holmium spins pointing inwards and two holmium spins pointing outwards. This is the so-called "ice rule", from the analogy with the H-O bond lengths in solid water. A local breaking of the ice rule, due to the flipping of one holmium spin shared between two neighbouring tetrahedra, results in a 3-in-1-out and a 1-in-3-out configuration. This effectively creates a positive and a negative magnetic charge in the adjacent tetrahedra[8], which can be regarded as a monopole-antimonopole pair[9-10]. Such pairs can dissociate and the individual monopoles can move away from each other by flipping a chain of spins along their route. This process takes place without further violating the ice rule, so that the energy cost for the monopoles to be brought to infinity stays finite, being linked to the energy required for the first excitation. Signatures of emergent magnetic monopoles in bulk spin ice crystals have been observed by several groups[11-15]. The magnetic counterparts of two fundamental effects in electronics have also been demonstrated: a basic capacitor effect for magnetic charges[16] and the introduction of magnetic defects hindering the monopole flow, similar to residual defect-induced resistance for electrons[17]. Further effort has been dedicated to tune the monopole chemical potential in a range where mutual interaction plays a role, to mimic electronic correlations in a purely magnetic Coulomb gas[18]. In conjunction with all the mentioned experiments, the term "magnetricity" is coined to describe the flow of magnetic charges as the equivalent of electricity.

However, until now all experiments on spin ice materials have been performed on bulk crystals. The possibility of making thin films of these materials will further advance the research into spin ice systems and the manipulation of the monopole states, and presents a crucial step towards the generation of real devices for "magnetronics".

We use pulsed laser deposition (PLD) to grow thin films of Ho$_2$Ti$_2$O$_7$ from a stoichiometric target. As substrates we use yttria-stabilised zirconia (YSZ) in the (001), (011) and (111) orientations. The lattice mismatch of YSZ with Ho$_2$Ti$_2$O$_7$ is 2% (Fig. 1a). The thickness of the deposited films is varied between 9 nm and 110 nm, as determined by X-ray reflectivity and cross-sectional scanning transmission electron microscopy (STEM). High-angle annular dark field STEM (HAADF-STEM) studies reveal that the films are single crystalline (Fig. 1b) and epitaxially grown on the YSZ (Fig. 1c).

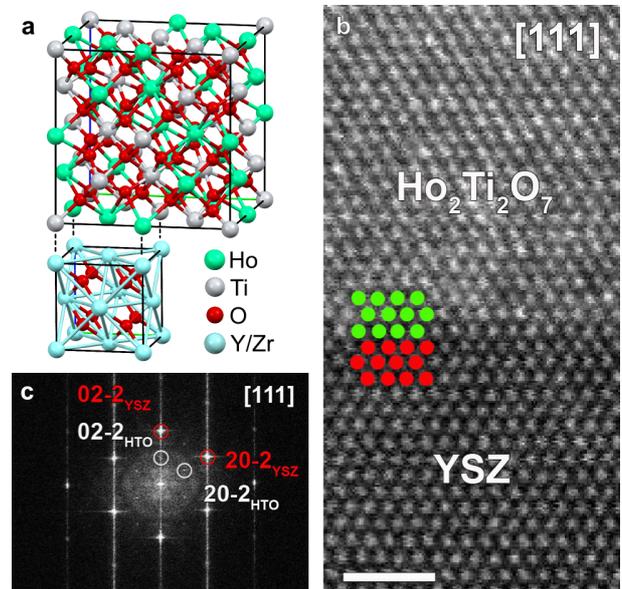

**Fig. 1.** (**a**) The pyrochlore structure of Ho$_2$Ti$_2$O$_7$ compared to the fluorite structure of YSZ. The lattice mismatch of Ho$_2$Ti$_2$O$_7$ with YSZ is 2%. (**b**) HAADF-STEM image of the (011)-oriented Ho$_2$Ti$_2$O$_7$/YSZ interface imaged along the [111] zone axis orientation. The Ho/Ti atomic columns in the Ho$_2$Ti$_2$O$_7$ layer are indicated by green dots, the Y/Zr atomic columns in the YSZ substrate by red dots. The indicated scale bar is 1 nm. (**c**) Fourier transform of the Ho$_2$Ti$_2$O$_7$ (HTO)/YSZ interface imaged along the [111] zone axis, evidencing the perfect epitaxial relationship of the layer with respect to the substrate.

The surface morphology of a (111)-oriented YSZ substrate prior to deposition, imaged by atomic force microscopy (AFM), is shown in Fig. 2a. Terraces with steps of 3.1 ± 0.1 Å are observed, which corresponds to one third of the lattice constant of YSZ along the (111) direction. Even after deposition of a $Ho_2Ti_2O_7$ film of 110 nm, this terrace structure is still visible (Fig. 2b). Fig. 2c shows an AFM image on a smaller scale. The typical peak-to-peak values of the films grown on differently oriented YSZ are between 1.5 – 5.5 nm and the root-mean-squared roughness of the films is between 0.4 – 1.2 nm. These low values indicate that the films are very smooth.

X-ray diffraction (XRD) is used to determine the quality of the crystal structure of the $Ho_2Ti_2O_7$ thin films. Fig. 3a shows symmetric $2\theta$-$\omega$ scans for ~110 nm $Ho_2Ti_2O_7$ films grown on differently oriented substrates. The presence of only out-of-plane peaks of $Ho_2Ti_2O_7$ further underlines the high quality epitaxial growth of the films. From these diffraction peaks a lattice constant of 10.1 Å is obtained, corresponding with the expected stoichiometry. We note that a few samples have a lattice constant up to 10.2 Å. This could indicate an off-stoichiometry of the Ho and Ti ratio for those films[19], possibly due to target inhomogeneity. The in-plane orientation of the crystal structure is investigated by performing $\phi$ scans around the (226) diffraction peak of $Ho_2Ti_2O_7$ (Fig. 3b). The expected four peaks, due to cubic symmetry, appear for the same $\phi$ angles as the (113) substrate diffraction peaks, confirming the cube-on-cube growth of the $Ho_2Ti_2O_7$ film on the substrate. Fig. 3c shows the reciprocal space map of a 33 nm thick $Ho_2Ti_2O_7$ film. The (113) YSZ and the (226) $Ho_2Ti_2O_7$ diffraction peaks appear at different in-plane values of the scattering vector $Q$. The film is therefore mainly relaxed and not fully strained to the substrate. This is expected for such a large thickness, but films with a thickness of 9 nm also show relaxation. This suggests that the elasticity of the crystal structure of $Ho_2Ti_2O_7$ is low. The presence of Kiessig fringes in Fig. 3c also underlines the smoothness of the grown film.

Although the studies on the crystal structure confirm the high quality of the $Ho_2Ti_2O_7$ films, the most important question is whether these films also show the spin ice behaviour. To verify this, the magnetic properties were studied by vibrating sample magnetometry (VSM). The investigated field directions are shown in Fig. 4a. $Ho_2Ti_2O_7$ films grown on differently oriented substrates are investigated to determine the in–plane magnetisation with different crystal orientations of $Ho_2Ti_2O_7$. Fig. 4b shows the field dependence of the magnetisation for a $Ho_2Ti_2O_7$ film grown on a (011)-oriented substrate and measured along the [111] direction at a temperature of 2 K, which is near to the onset of the spin correlations[20]. The absence of ferromagnetic hysteresis is expected due to the frustration in spin ice materials and the fact that the temperature of this measurement is slightly above the spin freezing temperature[3,21]. These observations are similar for all films grown on differently oriented YSZ substrates. The remarkable difference between the [111] direction and the other directions is the observation of an intermediate plateau at about 1.5 T. This is a hallmark for the spin ice behaviour[20,22-24]. For small magnetic fields applied along the [111] direction one of the holmium spins in the tetrahedron is not aligned along this field due to the ice rule. The step-like enhancement of the magnetisation appears when this spin aligns with the magnetic field and thus with breaking of the ice rule.

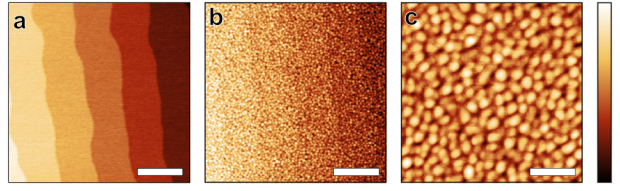

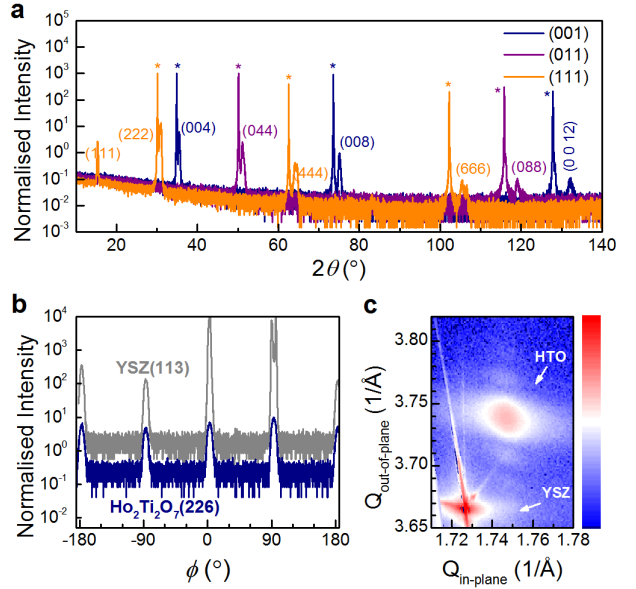

Fig. 2. Atomic force microscopy on a 110 nm $Ho_2Ti_2O_7$ film grown on (111)-oriented YSZ substrate. The colour scale on the right is between 0–7 nm and corresponds to the images in (b) and (c). (a) Image of (111)-oriented YSZ substrate prior to deposition. The step size corresponds to one third of the YSZ unit cell along the (111) direction. The indicated scale bar is 500 nm. (b) Image of a 110 nm $Ho_2Ti_2O_7$ film grown on the YSZ substrate shown in (a). The indicated scale bar is 500 nm. (c) Zoomed-in image of the $Ho_2Ti_2O_7$ film shown in (b). The indicated scale bar is 125 nm.

Fig. 3. X-ray diffraction studies on $Ho_2Ti_2O_7$ thin films. (a) Symmetric $2\theta$-$\omega$ scans of 110 nm thick $Ho_2Ti_2O_7$ films grown on differently oriented YSZ substrates. The diffraction peaks marked with an asterisk originate from the substrate. (b) $\phi$ scan of the (226) diffraction peak of a 110 nm $Ho_2Ti_2O_7$ film compared to the (113) diffraction peak of the (001)-oriented YSZ substrate. The graphs are displaced vertically for clarity. (c) Reciprocal space map of the (226) diffraction peak of a 33 nm thick $Ho_2Ti_2O_7$ (HTO) film grown on a (001)-oriented YSZ substrate. The colour scale is logarithmic and spans 5 decades.

To verify whether, besides this distinct feature for spin ice behaviour, the expected magnetic anisotropy[22-24] is also present in these thin films, we compare the saturation magnetisations along different crystal directions (Fig. 4c). The measured in-plane orientations are the [100] and [110] orientations for the (001)-oriented sample, and the [111] and [$\bar{1}2\bar{1}$] for the (011)-oriented sample. The expected averaged magnetisations per holmium atom are $\mu/\sqrt{3}$ for the [100] orientation, $\mu/2$ for the [111] orientation, and $\mu/\sqrt{6}$ for the [110] and the [$\bar{1}2\bar{1}$] orientations, where $\mu$ is the total moment of $10\mu_B$ for one holmium spin[22]. These values are indicated in Fig. 4c by dashed lines. To account for any thickness variations, we scale the measurements on the (001)-oriented sample relative to the expected value of the strongest signal, namely along the [100] direction. The measurements along the [$\bar{1}2\bar{1}$] direction of the (011)-orientated sample is scaled relative to measurements along the [110] direction of the (001)-oriented sample, as the [$\bar{1}2\bar{1}$] and [110] directions share the same expected value. We remark that the film thicknesses extracted from these scaled measurements correspond to the estimated thicknesses of 110-120 nm, as derived from the number of PLD pulses. The anisotropy due to the crystal field is clearly visible in Fig. 4c and the saturations approach the expected values of the spin ice model. The small deviations from the expected values can possibly be explained by edge effects in these thin films, and the effects of the magnetic anisotropy in case of a slight misalignment during the VSM measurements[23]. The [110] or [$\bar{1}2\bar{1}$] direction are unfavourable directions due to the strong anisotropy in Ho$_2$Ti$_2$O$_7$. It is therefore difficult to maintain the sample exactly along the [110] or [$\bar{1}2\bar{1}$] direction. In addition to this, on each tetrahedron along the [110] and [$\bar{1}2\bar{1}$] directions only two or three holmium spins, respectively, should contribute to the magnetisation (see Fig. 4a). If there is a misalignment, the remaining holmium spins can also be aligned to the magnetic field direction and contribute to the measured magnetisation. The smaller value for the [111] direction can also be explained by a slight misalignment. In the ideal case, all four holmium spins on a tetrahedron contribute to the magnetisation. Rotating the sample in-plane, the orientation changes from the [111] direction towards the [$\bar{1}2\bar{1}$] direction and the contribution of four holmium spins to the total magnetisation will decrease towards a contribution of only three holmium spins. Combining the clear anisotropy with the observation of the plateau in the magnetisation curve for the [111] direction, we conclude that the Ho$_2$Ti$_2$O$_7$ films grown by PLD exhibit spin ice behaviour.

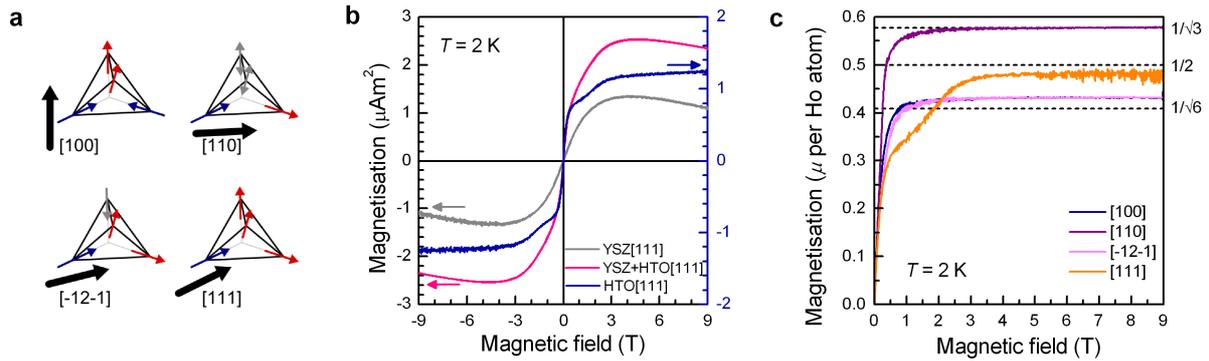

**Fig. 4. Magnetic characterisation of Ho$_2$Ti$_2$O$_7$ thin films.** (**a**) The investigated field directions (black arrows) compared to the orientation of the holmium spins on a tetrahedron. The grey holmium spins do not contribute to the magnetisation, as their orientation is perpendicular to the field. The blue spins point inwards the tetrahedron and the red spins point outwards. (**b**) Magnetisation along the [111] direction for an 80 nm Ho$_2$Ti$_2$O$_7$ film grown on (011)-oriented YSZ. The grey curve shows the magnetisation of the substrate, the pink curve is the magnetisation of the total sample and the blue curve is their difference, corresponding to the magnetisation of only the Ho$_2$Ti$_2$O$_7$ film. Note that the scale of the blue curve is given by the right axis. (**c**) Magnetisation of 110-120 nm Ho$_2$Ti$_2$O$_7$ films as function of applied field along different crystal orientations showing the magnetic anisotropy. Units of the magnetisation are in averaged magnetic moment per holmium atom $\mu$. The dotted lines indicate the theoretical values for the saturation magnetisations. The saturation magnetisation of the [100] orientation is scaled to the expected value of $\mu/\sqrt{3}$ and all other measurements are scaled relative to the saturation of the [100] orientation.

We have demonstrated the single crystalline thin film growth of the spin ice compound Ho$_2$Ti$_2$O$_7$, retaining the spin ice behaviour. The ability to grow spin ice thin films opens up new research directions on the spin ice dynamics in these frustrated systems. It is now possible to structure the spin ice material into devices and combine it with other materials by patterning on the micro- and nanoscale, e.g. to manipulate the monopoles via their electric dipoles using electrical gating[25]. The spin ice material itself can be easily modified by varying target composition or deposition conditions, or by post-annealing. This, and epitaxial strain engineering, may allow changing the chemical potential and the density of monopoles to enter the regime of monopole correlations[18]. We notice that also other pyrochlore oxides show, or are predicted to show, interesting properties. An example is Y$_2$Ir$_2$O$_7$, which is suggested to be a topological Mott insulator[26]. Also for those materials, the development of thin film growth will be of great interest, which may be encouraged by the example of Ho$_2$Ti$_2$O$_7$ presented here.

## Acknowledgements

The authors acknowledge support from the Dutch FOM and NWO foundations and from the European Union under the Framework 7 program under a contract from an Integrated Infrastructure Initiative (Reference 312483 ESTEEM2). G.V.T. acknowledges the ERC Grant N246791-COUNTATOMS. S.T. gratefully acknowledges financial support from the Fund for Scientific Research Flanders (FWO). The microscope used in this study was partially financed by the Hercules Foundation of the Flemish Government. The authors acknowledge fruitful interactions with A. Brinkman, M. G. Blamire, M. Egilmez, F. J. G. Roesthuis, J. N. Beukers, C. G. Molenaar, M. Veldhorst, and X. Renshaw Wang.

# Supplementary Information

# Thin films of the spin ice compound $Ho_2Ti_2O_7$

D. P. Leusink[1], F. Coneri[1], M. Hoek[1], S. Turner[2], H. Idrissi[2], G. Van Tendeloo[2], H. Hilgenkamp[1*]

[1]Faculty of Science and Technology & MESA+ Institute for Nanotechnology, University of Twente, P.O. Box 217, 7500 AE Enschede, The Netherlands. [2]Electron Microscopy for Materials Science (EMAT), Department of Physics, University of Antwerp, Groenenborgerlaan 171, B-2020 Antwerp, Belgium.
*e-mail: h.hilgenkamp@utwente.nl

**Methods**

A KrF excimer laser with a wavelength of 248 nm is used for the pulsed laser deposition. The repetition rate is set to 4 Hz and a laser fluency of 1.8 J/cm$^2$ with a spot size of 5.0 mm$^2$ is used on the stoichiometric target. This target is prepared using solid-state synthesis. High purity powders of $Ho_2O_3$ and $TiO_2$ are mixed and pressed into pellets. These pellets are sintered at 1200°C for a total of 36 hours, with intermediate milling and pressing steps. Prior to deposition the yttria-stabilised zirconia (YSZ) substrates are annealed at 1050°C in air to obtain a smooth surface. The substrates are mounted on a heater with silver paint and the deposition is done at a pressure of 0.15 mbar oxygen. Films grown in a temperature range of 750°C up to 900°C all show good crystal quality according to X-ray diffraction studies and the films described in this report are all grown at a heater temperature of 800°C. After deposition, the films are cooled to room temperature at a rate of 10°C/min at deposition pressure.

High-angle annular dark field scanning transmission electron microscopy (HAADF-STEM) experiments were carried out on a FEI Titan 80-300 "cubed" microscope fitted with an aberration-corrector for the probe forming lens operated at 300 kV, using a convergence semi-angle α of ~ 22 mrad and a collection semi-angle β of ~ 50 mrad. The samples for STEM investigation were prepared by focused ion beam (FIB) processing in a FEI Helios FIB. In addition to Fig. 1b, a HAADF-STEM image of the (011)-oriented $Ho_2Ti_2O_7$/YSZ interface of the complete $Ho_2Ti_2O_7$ film is shown in Fig. S1. Upon zooming in, the Ho/Ti atoms of the film and the Y/Zr atoms of the substrate are visible throughout the image.

The vibrating sample magnetometry (VSM) measurements were performed using a Quantum Design Physical Property Measurement System. For the magnetic characterisation of each sample, the substrate has been measured before the deposition of the film, so that the magnetic signal due to the film can be extracted. Furthermore, the backside of the substrate is polished to remove the silver paint after the $Ho_2Ti_2O_7$ deposition. All measurements are performed for positive and negative fields. To facilitate comparison between different directions for the same film, the sample has been rotated on the sample holder for the VSM.

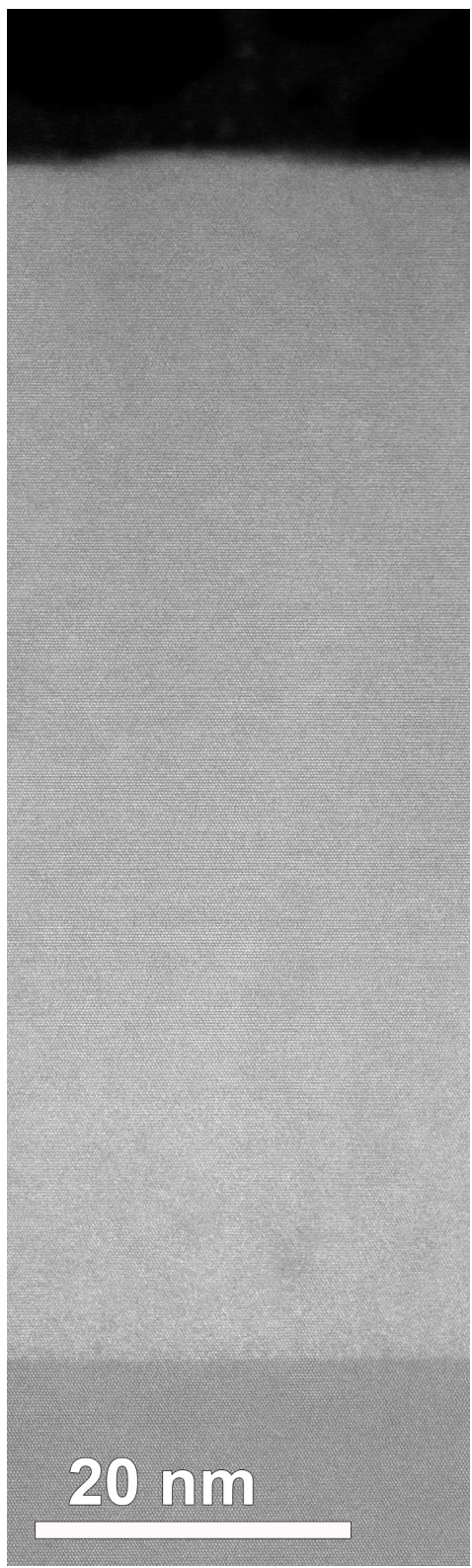

**Fig. S1.** HAADF-STEM image of the complete (011)-oriented $Ho_2Ti_2O_7$ film on YSZ along the [111] zone axis direction. The thickness of the film is 80 nm.